\documentclass[a4,12pt]{article}
\usepackage{graphicx}
\usepackage{amsmath}
\usepackage{amssymb}
\usepackage{rotating}
\usepackage[comma,authoryear]{natbib}

\DeclareMathOperator*{\argmin}{arg\,min}
\DeclareMathOperator*{\argmax}{arg\,max}

\newtheorem{lemma}{Lemma}
\newtheorem{proof}{Proof}
\newtheorem{theorem}{Theorem}
\newtheorem{algorithm}{Algorithm}
\begin{document}

\title{Tukey depth: linear programming and applications}

\author{Pavlo Mozharovskyi\\
{\small Centre Henri Lebesgue} \\
{\small Agrocampus Ouest} \\
{\small Institute of Mathematical Research of Rennes}\\
\indent\\
}

\date{February 29, 2016}

\maketitle

\begin{abstract}
Determining the representativeness of a point within a data cloud has recently become a desirable task in multivariate analysis.
The concept of statistical depth function, which reflects centrality of an arbitrary point, appears to be useful and has been studied intensively during the last decades.
Here the issue of exact computation of the classical Tukey data depth is addressed.
The paper suggests an algorithm that exploits connection between the Tukey depth and linear separability and is based on iterative application of linear programming.
The algorithm further develops the idea of the cone segmentation of the Euclidean space and allows for efficient implementation due to the special search structure.
The presentation is complemented by relationship to similar concepts and examples of application.
\end{abstract}


\section{Introduction}

Determining the representativeness of a point within a bunch of data or a probability measure has recently become a desirable task in multivariate analysis. Nowadays it finds applications in different domains of economics, biology, geography, medicine, cosmology and many others. In his celebrated work, \citet{Tukey75} introduced an idea to order multivariate data, which has later been developed by \citet{DonohoG92} and is known as the \emph{Tukey} (=\emph{halfspace}, \emph{location}) \emph{depth}. Generally, the statistical data depth is a function determining how centrally a point is located in a data cloud. The upper-level sets it generates --- trimmed regions --- are set-valued statistics. They trim data w.r.t. the degree of centrality. For more information on the data depth the reader is referred to \citet{ZuoS00,Dyckerhoff04,Mosler13} and mentioned there references.

The Tukey data depth is one of the most important depth notions and is historically the first one. Regard a random vector $X$ distributed as $P$, in particular empirically on $\{\mathbf{x}_1,...,\mathbf{x}_n\}$, in $\mathbb{R}^d$. The Tukey depth of a point $\mathbf{z}\in\mathbb{R}^d$ w.r.t. $X$, further $D({\mathbf z}|X)$, is defined as the smallest probability mass of a closed halfspace containing $\mathbf{z}$:
\begin{equation}
    D({\mathbf z}|P)=\inf\{P(H)\,|\,H\mbox{ closed half-space}, \mathbf{z}\in H\}.
\end{equation}

The Tukey depth possesses many desirable properties: it is affine invariant, tends to zero at infinity, is monotone on rays from any deepest point, quasiconcave and upper semicontinuous. By that it satisfies all the postulates imposed on a depth function~\citep{ZuoS00,Dyckerhoff04,Mosler13}. If $P$ is absolutely continuous, the Tukey depth is a continuous function of ${\mathbf z}$ achieving maximum value of $\frac{1}{2}$, for angularly symmetric distributions at the center of symmetry. If $P$ has no Lebesgue density, the Tukey depth is a discrete function of ${\mathbf z}$ and can have a non-unique maximum. By definition, its empirical version vanishes beyond the convex hull of the data. The Tukey depth determines uniquely empirical distribution~\citep{Koshevoy02}, taking a finite number of values in the interval from 0 (for the points lying outside the convex hull of the data) to $\frac{1}{2}$, increasing by a multiple of $\frac{1}{n}$. Naturally, it has attractive breakdown properties and converges for a sample from $P$ almost surely to the depth w.r.t. $P$~\citep{DonohoG92}.

For a data cloud $D({\mathbf z}|X)$ can be expressed as the smallest portion of $X$ to be cut off by a hyperplane through $\mathbf{z}$ so that the remaining points lie in an open halfspace not containing $\mathbf{z}$:
\begin{equation}\label{eqn:TukeyDepth}
    D({\mathbf z}|X)=\frac{1}{n}\min_{{\mathbf r}\in S^{d-1}}\#\{i|{\mathbf x}_i^\prime{\mathbf r}\ge{\mathbf z}^\prime{\mathbf r},{\mathbf x}_i\in X\}.
\end{equation}

Exact calculation of the Tukey depth is a computationally challenging task of non-polynomial complexity. For this reason, great part of the literature on the Tukey depth concerns its computational aspects. The reader is referred to \citet{DyckerhoffM15} for the exact algorithm and a reference to some preceding works. \citet{Dyckerhoff04} introduced the weak projection property, which is satisfied inter alia by the Tukey depth. This allows to approximate the depth as the minimum over univarite depths in the projections onto one-dimensional spaces. For the latest research in this area see~\citet{ChenMW13} and contained there references.

In the current paper an algorithm based on linear programming is proposed. Here, the idea of the conic segmentation of $\mathbb{R}^d$, introduced by~\citet{MoslerLB09} for constructing zonoid trimmed region, is exploited. Applied to the Tukey depth this can be formulated as follows: The entire space is divided into polyhedral cones, each having --- in the projection onto any direction in its interior --- the same subset of $X$ above (below) the projection of $\mathbf{z}$, and thus delivering the same univariate Tukey depth. For each of the cones this depth value is calculated, and the Tukey depth is then the minimum over all these depths. Then, starting from an arbitrary cone, all cones are regarded by means of the breadth-first search algorithm. A procedure employing this principle for computation of the Tukey depth has been suggested by \citet{LiuZ14a}, where the convex hull algorithm~\citep{BarberDH96} is used to detect neighboring ones.

In the proposed method, the connection between the problem of linear separability in binary classification problem and the Tukey depth is exploited. Based on this, linear programming is used for finding cone's neighbors, and each cone is coded by a binary sequence. First, this gives possibility to examine only candidates of interest for the neighboring cones, separately.
Second, the number of the candidates to be checked can be substantially reduced. Third, when employing the binary coding, one does not need to compute the cone's location in $\mathbb{R}^d$, which allows for reducing precision problems and further saves computational expenses. Also, the calculations are performed in the spaces of dimension $d-1$, by a simplex algorithm. Finally, linear programming allows for caching by remembering (last) found basis for each hyperplane.

To be precise, below the following task is being solved: Given a data sample $X=\{\mathbf{x}_1,...,\mathbf{x}_n\}\in\mathbb{R}^d$, $d<n$, and a point $\mathbf{z}\in\mathbb{R}^d$, the Tukey depth of ${\mathbf z}$ w.r.t. $X$ shall be calculated. For the rest of the paper, we assume that $\{\mathbf{z}\}\cup X$ are in general position, i.e., every subset of $k+1$ points $\in(\{\mathbf{z}\}\cup X)$ spans a subspace of dimension $\min\{k,d\}$. Violation of these assumptions can be compensated by a location shift and a slight perturbation of the data. The Tukey depth is discrete, so such a perturbation can be potentially harmful, as only a small shift of one point can change the depth value of $\mathbf{z}$ in a non-continuous way. Before performing such a perturbation, we suggest to first check whether $\mathbf{z}\in\mbox{conv}(X)$ (if not, $D(\mathbf{z}|X)=0$), and only then calculate the depth of $\mathbf{z}$ using perturbed data. When $n$ is not very small and the zero-depth case is specially treated, possible perturbation damage is negligible.

The rest of the paper is organized as follows.
First, in Section~\ref{sec:connections}, we regard connection of the concept of the Tukey depth to some important problems of the multivariate analysis.
Then, Section~\ref{sec:theory} provides the theoretical results for the proposed algorithm, which is given in Section~\ref{sec:algorithm}.
Further, Section~\ref{sec:applications} suggests two demonstrative examples where application of the Tukey depth can be advantageous.
Section~\ref{sec:outlook} concludes.

\section{Connections}\label{sec:connections}

Due to its indicator-loss nature, the Tukey depth is known to be connected to a number of problems. Although just the smallest portion of the points to be cut off by a closed halfspace is required as an output, in many algorithms the boundary hyperplane is found or can be restored based on the output information. For shortness, let $R=\argmin_{{\mathbf r}\in S^{d-1}}\#\{i|{\mathbf x}_i^\prime{\mathbf r}\ge{\mathbf z}^\prime{\mathbf r},{\mathbf x}_i\in X\}$ be the set of directions $\mathbf{r}\subset S^{d-1}$ each achieving $\frac{1}{n}\#\{i|{\mathbf x}_i^\prime{\mathbf r}\ge{\mathbf z}^\prime{\mathbf r},{\mathbf x}_i\in X\}=D(\mathbf{z}|X)$.

\emph{Densest hemisphere}. The problem of computing the Tukey depth is invariant more than just in the affine way. Thus shifting $X$ to get $\mathbf{z}$ in the origin and projecting $X$ onto the unit sphere $S^{d-1}$ after that (a non-affine transformation), changes neither the value of the depth $D({\mathbf z}|X)$ nor the set of optimal normals to separating hyperplanes $R$, \emph{i.e.} $D({\mathbf z}|X)$ = $D(\mathbf{0}|Y)$ as well as the optimal argument set with $Y=\{\frac{\mathbf{x}_i - \mathbf{z}}{\|\mathbf{x}_i - \mathbf{z}\|}|i=1,...,n\}\subset S^{d-1}$. The last one corresponds to the \emph{open densest hemisphere} problem shown by \citet{JohnsonP78} to be of non-polynomial time complexity, namely $O(n^{d-1}\log n)$ if dimension $d$ is fixed, for the Tukey depth, see \citet{DyckerhoffM15}.

\emph{Classification}. By a trivial modification, the task of \emph{supervised binary linear classification} can be narrowed down to finding an optimal argument $\mathbf{r}$ from (\ref{eqn:TukeyDepth}), see also \citet{GhoshC05a}. Indeed, regard $X_1=\{\mathbf{x}_1,...,\mathbf{x}_{n_1}\}$ and $X_2=\{\mathbf{x}_{n_1+1},...,\mathbf{x}_{n_1+n_2}\}$ being two training classes in $\mathbb{R}^d$, and let $Y=\{\mathbf{x}_i - \mathbf{x}_j|i=1,...,n_1,\,j=n_1+1,...,n_1+n_2\}$. When minimizing empirical risk of a linear classifier, we are interested in a direction $\mathbf{r}\in S^{d-1}$ in projection on which possibly many differences $\mathbf{x}_i - \mathbf{x}_j$ have the same sign, or in other words, as many (few) as possible points from $Y$ lie on the same side of a hyperplane through $\mathbf{0}$. The last holds for any element of $R$.

\emph{Regression depth}. \citet{RousseeuwH99} define data depth for a liner regression model based on the notion of nonfit: Given regression input $X=\{\mathbf{x}_1,...,\mathbf{x}_n\}$ in $\mathbb{R}^d$ and output $\mathbf{y}=\{y_1,...,y_n\}$ in $\mathbb{R}$, a fit $\mathbf{b}=(b_0,b_1,...,b_d)$ is called a nonfit if there exists an affine hyperplane in the input space not containing any point from $X$ such that all the points from $X$ lying on its same side have residuals strictly of the same sign. \emph{Regression depth} of a fit $\mathbf{b}\in\mathbb{R}^{d+1}$ is the smallest number of observations to be removed from $X$ sufficient for $\mathbf{b}$ to become a nonfit. Given a fit $\mathbf{b}\in\mathbb{R}^{d+1}$, after splitting $X$ into two classes on the basis of the sign of residuals (and acting as in the preceding paragraph), regression depth is given by the empirical risk in the projection on any element of $R$.

\emph{Maximum feasible subsystem}. Again, let $Y=\{y_i=\frac{\mathbf{x}_i - \mathbf{z}}{\|\mathbf{x}_i - \mathbf{z}\|}|i=1,...,n\}\subset S^{d-1}$. First consider the case $d = 2$, where $Y$ lies on the unit circle. Here $y_i$ defines an open halfcircle, in which each point defines a closed halfspace with the origin on its boundary and never containing $y_i$. Then the task of computation of the Tukey depth narrows down to finding a point on the unit circle contained in the largest number of these halfcircles; the set of such points coincides with $R$. Extending this logic to higher dimensions \citep{BremnerCILM08} leads to (two instances of) the \emph{maximum feasible subsystem} problem in $\mathbb{R}^{d-1}$. For $d>2$, in the same way, each $y_i$ defines an open halfspace in which each element (=point $\in\mathbb{R}^d$) is a normal to a hyperplane defining a halfspace with the origin on its boundary and not containing $y_i$. We are interested in a point lying in the intersection of the highes number of these halfspaces. Let $H_+=\{(x_1,...,x_d)^\prime|x_d = 1\}$ ($H_-=\{(x_1,...,x_d)^\prime|x_d = -1\}$) be a positive (respectively negative) hyperplane. For each $y_i$, intersection of such open halfspace with $H_+$ yields an open halfspace in $H_+$ of dimension $\mathbb{R}^{d-1}$; the same holds for $H_-$. Then, the maximum number such halfspaces either in $H_+$ or in $H_-$ having nonempty intersection divided through $n$ equals the Tukey depth. The two maximum feasible subsystem problems consist of linear inequalities $\mathbf{x}\frac{\mathbf{y}_{i,(1,...,d-1)}}{\|\mathbf{y}_{i,(1,...,d-1)}\|}\le -\tan(\frac{\pi}{2}-\arccos\mathbf{y}_{i,(d)})$ for $H_+$ and or $\mathbf{x}\frac{\mathbf{y}_{i,(1,...,d-1)}}{\|\mathbf{y}_{i,(1,...,d-1)}\|}\le \tan(\frac{\pi}{2}-\arccos\mathbf{y}_{i,(d)})$ for $H_-$. (Note that due to the general position assumption there is no difference between open and closed halfspaces, and the equator $\{(x_1,...,x_d)^\prime|x_d = 0\}$ should not be searched through because $R$ has nonzero volume on $S^{d-1}$.)

\cite{HallinPS10} establish connection of the Tukey depth to the linear programming via quantile regression \citep[see also][]{KoenkerB78} and exploit this to compute Tukey depth regions. In what follows, connection between the Tukey depth and linear programming via linear separability is used as a basis for construction of an algorithm computing the Tukey depth.

\section{Theoretical background}\label{sec:theory}

\begin{figure}[t!]
\begin{center}
    \includegraphics[keepaspectratio=true,scale=0.25]{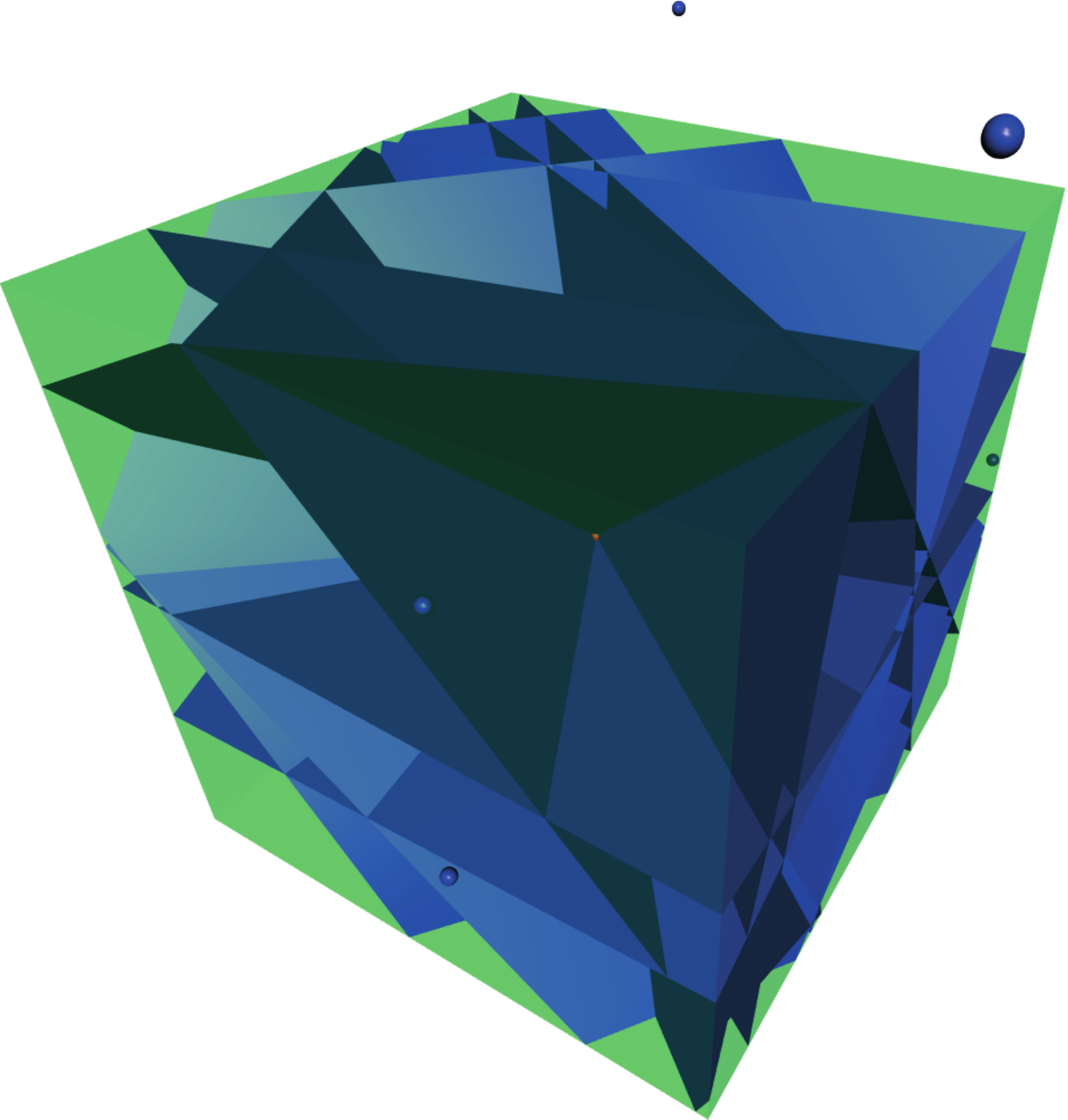}\quad
	\includegraphics[keepaspectratio=true,scale=0.9]{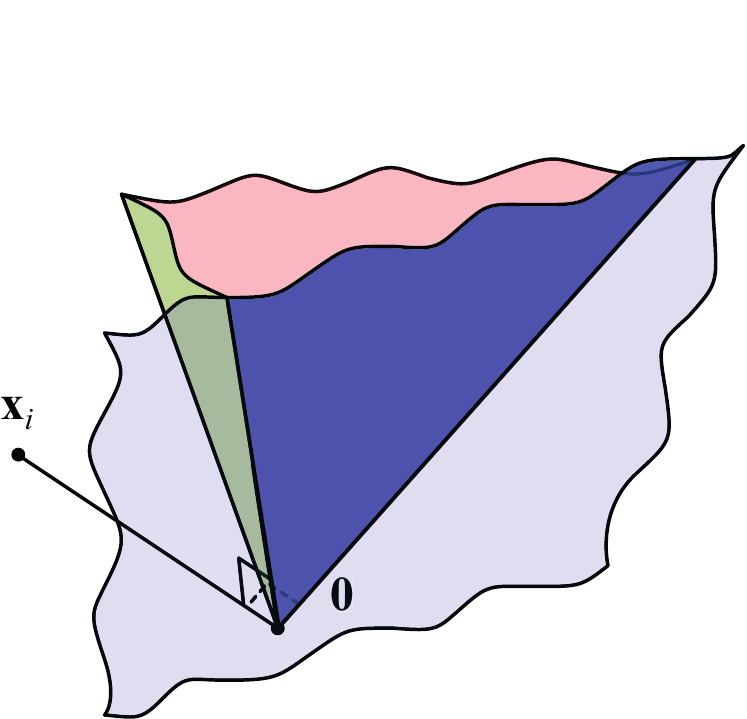}
	\caption{Cone segmentation on the unit cube (left) and a cone's facet defined by a point (right)}
    \label{fig:pic1}
\end{center}
\end{figure}

For simplicity and w.l.o.g., assume for the rest that $\mathbf{z}=\mathbf{0}$. Consider a direction, i.e. a point on the unit sphere $\mathbf{r}\in S^{d-1}$. It yields an ordered sequence, a permutation $\pi_{\mathbf{r}}$ on ${\mathcal N}=\{1,...,n\}$ such that $\mathbf{x}_{\pi_{\mathbf{r}}(1)}^\prime\mathbf{r}\le\mathbf{x}_{\pi_{\mathbf{r}}(2)}^\prime\mathbf{r}\le...\le\mathbf{x}_{\pi_{\mathbf{r}}(n)}^\prime\mathbf{r}$. If the data are in general position a vector $\mathbf{r}$ can be found such that all inequalities hold strictly $\mathbf{x}_{\pi_{\mathbf{r}}(1)}^\prime\mathbf{r} < \mathbf{x}_{\pi_{\mathbf{r}}(2)}^\prime\mathbf{r} < ... < \mathbf{x}_{\pi_{\mathbf{r}}(n)}^\prime\mathbf{r}$, and $\mathbf{x}_{\pi_{\mathbf{r}}(i)}^\prime\mathbf{r}\ne0,i=1,...,n$. Then such $\mathbf{r}$ splits $X$ into two disjoint subsets (by its normal hyperplane $H_{\mathbf{r}}$ through $\mathbf{0}$ yielding two open halfspaces $H_{\mathbf{r}}^+$ and $H_{\mathbf{r}}^-$ in ${\mathbb R}^d$), $X_{\mathbf{r}}^+=\{\mathbf{x}\in X|\mathbf{x}^\prime\mathbf{r}>0\}$ and $X_{\mathbf{r}}^-=\{\mathbf{x}\in X|\mathbf{x}^\prime\mathbf{r}<0\}$ containing the points with strictly positive, respectively negative, projections on $\mathbf{r}$. Let us call the closure of the set of all $\lambda\mathbf{r},\lambda\ge0$, maintaining the same $X_{\mathbf{r}}^+$ and $X_{\mathbf{r}}^-$, a \emph{direction cone} $C$ (yielding $X_C^+=\{\mathbf{x}\in X|\mathbf{x}^\prime\mathbf{r}>0\,\forall\,\mathbf{r}\in\mbox{int}(C)\}$ and $X_C^-=\{\mathbf{x}\in X|\mathbf{x}^\prime\mathbf{r}<0\,\forall\,\mathbf{r}\in\mbox{int}(C)\}$ respectively). This is because its form constitutes an infinite polyhedral cone with the apex in the origin. The entire $\mathbb{R}^d$ is then filled by the set of all direction cones, say ${\mathcal C}(X)$, while each cone $C\in{\mathcal C}(X)$ defines some portion of the sample, that can be cut off by the hyperplane normal to any $\mathbf{r}\in\mbox{int}(C)$. Denote this portion $D_C(\mathbf{0}|X)=\frac{1}{n}\min\{\sharp X_C^+,\sharp X_C^-\}$ ($\sharp$ stands for the set's cardinality), then the Tukey depth is $D(\mathbf{0}|X)=\min_{C\in {\mathcal C}(X)}{D_C(\mathbf{0}|X)}$. Below ${\mathcal C}(X)$ will be mentioned as \emph{cone segmentation}; see Figure~\ref{fig:pic1} left for a cone segmentation on the unit cube for ten standard normal deviates. One of the direction cones can be seen in the unit cube's corner directed to the reader.

The further task is then to go through all such cones and to find the one(s) delivering the smallest $\frac{1}{n}\min\{\sharp X_C^+,\sharp X_C^-\}$, i.e., the Tukey depth. Starting with \citet{MoslerLB09}, the usual way to proceed is:
\begin{enumerate}
    \item [(1)] choose an arbitrary direction cone,
    \item [(2)] move from each direction cone to the neighbors,
    \item [(3)] by that cover the entire ${\mathbb R}^d$ using breath-first search algorithm,
    \item [(4)] on each step check whether a direction cone has already been considered, i.e. saved in a structure maintaining fast search (usually a binary search tree).
\end{enumerate}

Ad (1), the task is trivial: a direction $\mathbf{r}\in S^{d-1}$ maintaining the ordering with strict inequalities $\mathbf{x}_{\pi_{\mathbf{r}}(1)}^\prime\mathbf{r} < \mathbf{x}_{\pi_{\mathbf{r}}(2)}^\prime\mathbf{r} < ... < \mathbf{x}_{\pi_{\mathbf{r}}(n)}^\prime\mathbf{r}$ and no projection coinciding with $\mathbf{z}^\prime\mathbf{r}=0$ has to be generated. When drawing $\mathbf{r}$ randomly, the theoretical probability of this event $=1$. As in practice draw concerns only a finite number of digits, it can (though extremely rarely) happen that one needs more than one drawing.

\subsection{Identification of neighboring cones}\label{ssec:cones}

Ad (2), identifying neighboring direction cones (2a) and transition to each of them if new (2b) is to be done. Let us take a closer look at the direction cone. Two different cones $C_1$ and $C_2$ differ in their corresponding set pairs $(X_{C_1}^+,X_{C_1}^-)$ and $(X_{C_2}^+,X_{C_2}^-)$. So, if a point $\mathbf{r}\in S^{d-1}$ moves from one direction cone to another, projections of one or more points on $\mathbf{r}$ \emph{migrate passing the origin}, i.e., change the sign. Let $C_1$ and $C_2$ be two cones, such that a direct (i.e., not crossing other cones) rotational movement of $\mathbf{r}$ from $C_1$ to $C_2$ (and vice-versa) is possible. That means that $C_1$ and $C_2$ have an intersection of affine dimension between $1$ and $d-1$. If transition of $\mathbf{r}$ from $C_1$ to $C_2$ involves changing the halfspace (from $H_{\mathbf{r}}^+$ to $H_{\mathbf{r}}^-$ or vice versa) by one point $\in X$ only (correspondingly changing the sign of its projection on $\mathbf{r}$), then $C_1$ and $C_2$ intersect in affine dimension $d-1$. This intersection constitutes the cones' common facet. We call such two cones {\em neighboring cones}.

\begin{figure}[t!]
\begin{center}
	\includegraphics[keepaspectratio=true,scale=0.67]{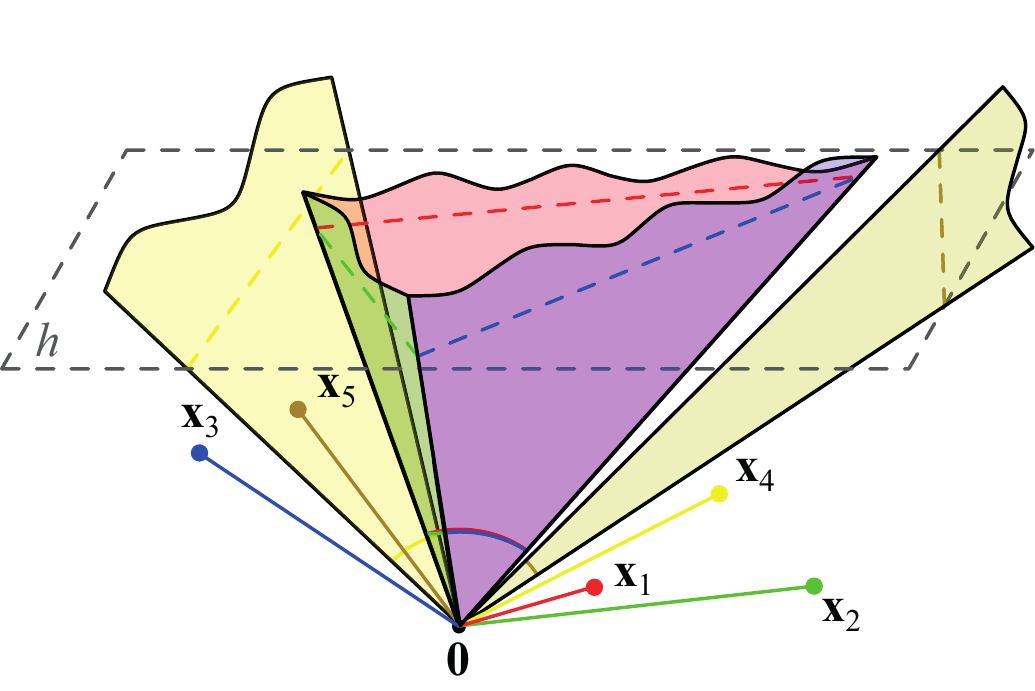}\quad
    \includegraphics[keepaspectratio=true,scale=0.67]{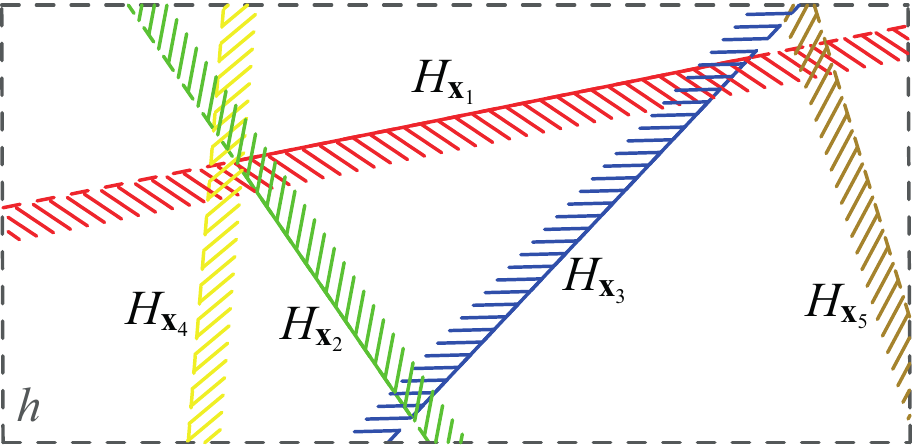}
	\caption{A direction cone in ${\mathbb R}^3$ defined by the points $\mathbf{x}_1$, $\mathbf{x}_2$ and $\mathbf{x}_3$, halfspaces formed by $\mathbf{x}_4$ and $\mathbf{x}_5$ are not directly involved (left); arbitrary cutting hyperplane $h$ visualizing how the hyperplanes are involved (right).}
    \label{fig:pic2}
\end{center}
\end{figure}

So, the transition of a single point $\mathbf{x}_i\in X$ from $X_{C_1}^+$ to $X_{C_2}^-$ means traversing of $\mathbf{r}$ from $C_1$ to a neighboring cone $C_2$ through a facet, and thus the facet is defined by this point $\mathbf{x}_i$, see Figure~\ref{fig:pic1} right. Naturally, given a cone $C$, any facet of $C$ lies in a hyperplane, normal to the line, connecting a point $\in X$ with $\mathbf{z}=\mathbf{0}$, as it is shown in Figure~\ref{fig:pic1} right, but not each point $\in X$ generates a facet of $C$, see Figure~\ref{fig:pic2}. A direction cone $C$ is defined by the intersection of closed halfspaces $\{\mathbf{y}|\mathbf{y}^\prime(\mathbf{x}-\mathbf{z})\ge 0,\mathbf{x}\in X^+_C\}$ and $\{\mathbf{y}|\mathbf{y}^\prime(\mathbf{x}-\mathbf{z})\le 0,\mathbf{x}\in X^-_C\}$. Hyperplanes directly involved in the intersection (generated by points $\mathbf{x}_1,\mathbf{x}_2,\mathbf{x}_3$ in Figure~\ref{fig:pic2}) contain the cone's facets and those outside (generated by points $\mathbf{x}_4,\mathbf{x}_5$ in Figure~\ref{fig:pic2}) do not. Thus, given a direction cone, a natural question is: Which points $\in X$ define its facets, and which do not? This is summarized in Theorem~\ref{thm:theorem1}.

\begin{theorem}
\label{thm:theorem1}
Given $X=\{\mathbf{x}_1,...,\mathbf{x}_n\}\in{\mathbb R}^d$, assume that $\{\mathbf{0}\}\cup X$ are in general position, and let $C$ be a direction cone. Also, for a point $\mathbf{x}\in X$ let $X_{H_{\mathbf{x}}}$ be the orthogonal projection of $X$ onto the $(d-1)$-dimensional linear subspace $H_{\mathbf{x}}$ normal to $\mathbf{x}$, and $X_{H_{\mathbf{x}},C}^+$ and $X_{H_{\mathbf{x}},C}^-$ be the two subsets of $X_{H_{\mathbf{x}}}\setminus\{{\mathbf 0}\}$ corresponding to $X_C^+$ and $X_C^-$, respectively. Then:
\begin{itemize}
    \item[(i)] $H_{\mathbf{x}}$ contains a facet of $C$ if and only if $X_{H_{\mathbf{x}},C}^+$ and $X_{H_{\mathbf{x}},C}^-$ are linearly strictly separable through $\mathbf{0}\in\mathbb{R}^{d-1}$, i.e., can be separated by a (d-2)-hyperplane $\subset H_{\mathbf{x}}$ containing the origin and no points from $X_{H_{\mathbf{x}},C}^+ \cup X_{H_{\mathbf{x}},C}^-$,
    \item[(ii)] if $\mathbf{r}\in S^{d-1}$ moves from $C$ to a neighboring direction cone through a facet $\subset H_{\mathbf{x}}$,
    the projection of ${\mathbf x}$ only on the line through ${\mathbf r}$ changes sign.
\end{itemize}
\end{theorem}

\begin{proof}
    \begin{itemize}
        \item [(i)] ``$\Longrightarrow$'': If $\mathbf{x}\in X$ defines a facet of $C$, then $H_{\mathbf{x}}$ contains this facet, and thus there should exist some direction $\mathbf{v}\in S^{d-1}\cap H_{\mathbf{x}}$ such that $X_C^+$ and $X_C^-$ projected onto $\mathbf{v}$ maintain their signs, except for the single point $\mathbf{x}$ being projected into $\mathbf{0}\in\mathbb{R}^{d-1}$.
            So, these projections of $X_C^+\setminus \{\mathbf{x}\}$ and $X_C^-$ (or $X_C^+$ and $X_C^-\setminus \{\mathbf{x}\}$) are separated in $H_{\mathbf{x}}$ by the hyperplane normal to $\mathbf{v}$ through $\mathbf{0}$.

            ``$\Longleftarrow$'': Strict linear separability of $X_{H_{\mathbf{x}},C}^+$ and $X_{H_{\mathbf{x}},C}^-$ through $\mathbf{0}$ means that there exists some $\mathbf{v}\in S^{d-1}\cap H_{\mathbf{x}}$, such that $\mathbf{x}^\prime\mathbf{v}>0$ $\forall$ $\mathbf{x}\in X_{H_{\mathbf{x}},C}^+$ and $\mathbf{x}^\prime\mathbf{v}<0$ $\forall$ $\mathbf{x}\in X_{H_{\mathbf{x}},C}^-$. Then a slight infinitesimal rotation of $\mathbf{v}$ towards (and inside of) the cone does not cause the projections to change signs, and thus maintains $X_C^+$ and $X_C^-$.
        \item [(ii)] Let $\mathbf{x}$ define a common facet of $C$ and $C^\prime$. As $\mathbf{x}$ is projected into $\mathbf{0}\in\mathbb{R}^{d-1}$ for all $\mathbf{r}\in S^{d-1}\cap H_{\mathbf{x}}$, then obviously when (slightly) deviating $\mathbf{v}$ to different sides of $H_{\mathbf{x}}$, the signs of $\mathbf{x}^\prime\mathbf{v}$ will be opposite. All points $\in\{\lambda\mathbf{x},\lambda\in\mathbf{R}\}$ change sign in their projection on $\mathbf{v}$, but as $\{\mathbf{0}\}\cup X$ are in general position, $\mathbf{x}$ is the only one.
    \end{itemize}
\end{proof}

\subsection{Optimization of the breadth-first search algorithm}

In Section~\ref{ssec:cones} we have addressed (2a) and (2b) by Theorem~\ref{thm:theorem1}. From the first part, one can easily find out which points define the cone's facets. Then, following the second part, moving the direction $\mathbf{r}$ to a neighboring cone by traversing their common facet constitutes in changing sign of the projection on $\mathbf{r}$ of the point which defines this facet.

Ad (3), we use the results from above to describe the {\em breadth-first search} algorithm: generate an initial direction cone (ad (1)) and move to the neighboring cones (ad (2)), calculating the depth in each of them, till the entire ${\mathbb R}^d$ is covered. Note, that covering a cone segmentation of $\mathbb{R}^d$ by a breadth-first search is general for some depth-calculating algorithms (\citet{LiuZ14a,LiuZ14b}) and algorithms constructing trimmed regions (\citet{MoslerLB09,PaindaveineS12a,PaindaveineS12b,BazovkinM12}) when $d\ge 3$. Below the algorithm is summarized to be referenced it in further explanations. The algorithms of \citet{PaindaveineS12a,PaindaveineS12b,LiuZ14a,LiuZ14b} differ from this one in that they store cones' facets and not cones while employing the convex hull algorithm.

The \emph{breadth-first search algorithm} on a cone segmentation of ${\mathbb R}^d$ proceeds in following steps:
\begin{itemize}
    \item[(a)] Draw an initial cone and store it in a queue.
    \item[(b)] Pop one cone from the head of the queue, process it, remember it, and for each of its neighboring cones do:\label{itm:2}
    \begin{itemize}
        \item[(c)] If the cone has not been processed till now push it into the tail of the queue.\label{itm:3}
    \end{itemize}
    \item[(d)] If the queue is not empty, go to Step (b).
\end{itemize}

Further, let us introduce the notion of the cone's {\em generation}, a number given to each cone (in Step \emph{c}) when it is pushed into the queue. The initially drawn cone (in Step \emph{a}) is given the initial number, say 1. (The generation can be thought of as the `depth' of the current searching path of the algorithm.) Obviously, when covering a cone segmentation of ${\mathbb R}^d$ with the breadth-first search algorithm, for processing cones of the $i$-generation, only cones of the $(i-1)$-, $i$- and $(i+1)$-generation have to be remembered. While on starting (low) generations the number of the cones from one generation to another grows rapidly, on close to `equatorial' generations (these basically constitute the segmentation) the increase is much less. Also, though the store-search structure for the cones is usually a binary tree, the computational time for search can be saved either, especially when the search is frequently performed.

Ad (4): When calculating the Tukey depth under the general position assumption of $\{\mathbf{z}\}\cup X$, one can step much further in this direction, and save less than tree generations. First, to simplify further presentation, let us code the cones. As mentioned above, the interior of each cone $C$ maintains the disjoint division of $X$ into $X_C^+$ and $X_C^-$ according to the signs in $X$'s projection onto any $\mathbf{r}\in\mbox{int}(C)$, and thus is uniquely defined by this division. So, binary identifiers for the cones can be used: a cone is coded by a binary sequence (``0'' and ``1'' say) of length $n$, where each bit represents a point $\in X$ w.r.t. some initial ordering of the points $\in X$ that is kept constant during the entire procedure. Points belonging to $X_C^+$ are coded by ``1'', those belonging to $X_C^-$ by ``0''.

After coding the initial cone ($C_0$ say) this way, other cones can be coded either the same way, or by another binary sequence identifying whether a point has changed the sign w.r.t. $C_0$ (``1'') or not (``0''). Then any cone's code can be obtained as the code of $C_0$ with those bits inverted that have been switched on in this sequence. This leads to Lemma~\ref{thm:lemma3}.

\begin{lemma}
    \label{thm:lemma3}
    Let us start the breadth-first search algorithm with an arbitrary initial cone $C_0$, and in Step~b, when checking for neighboring cones, always regard only cones defined by points which have not changed their sign in the projection onto $\mathbf{r}\in\mbox{int}(C_0)$ yet. Then in processing cones of the $i$-th generation, only cones of the $i$-th generation have to be remembered to check for neighboring cones and of the $(i+1)$-th generation to check whether a new cone has already been seen.
\end{lemma}

\begin{proof}
    If the cones defined by already processed points, i.e. those having changed their sign in the projection, are not considered, then only cones of the $(i+1)$-th generation can be taken into account when deciding whether a cone has already been seen. No cones of the $(i-1)$-th or $i$-th generation can be found because points defining them are not checked at all.
    Then one can go through all the cones of the $i$-th generation, and add those newly found from the $(i+1)$-th generation to the queue.
\end{proof}

Theorem~\ref{thm:theorem1} (ii) and Lemma~\ref{thm:lemma3} lead to Lemma~\ref{thm:lemma4}. Note that $\lfloor u\rfloor$ stands for the largest integer $\le u$.

\begin{lemma}
    \label{thm:lemma4}
    When starting the breadth-first search algorithm with an arbitrary initial cone $C_0$, and in Step~b regard only neighboring cones defined by points which have not changed their sign in the projection onto $\mathbf{r}\in\mbox{int}(C_0)$ yet, only $\lfloor\frac{n+2}{2}\rfloor$ generations have to be considered.
\end{lemma}

\begin{proof}
    From Theorem~\ref{thm:theorem1} (ii), each point may define a cone's facet, changing its sign in projections on all directions of the neighboring cone. If, following Lemma~\ref{thm:lemma3}, on each new step only not yet considered points are taken into account, then in each new generation exactly one point more has its sign on projection changed (compared to $C_0$). The maximum generation (if $C_0$ is denoted as $1$st generation) is then $(n+1)$-th generation.

    Each cone has its mirror-copy cone, where projections of $X$ on all directions have exactly opposite signs; these cones need not be considered, of course. Then, if $n$ is odd, exactly $\frac{n+1}{2}$ generations have to be considered, if $n$ is even, $\lfloor\frac{n+1}{2}\rfloor$ $+1$ generations have to be considered, as the mirror-copy cones of the `equatorial' (having number $\lfloor\frac{n+1}{2}\rfloor+1$) generation also belong to the equatorial generation. Thus, at most $\lfloor\frac{n+2}{2}\rfloor$ generations have to be regarded.
\end{proof}

\section{Algorithm}\label{sec:algorithm}

Basically, Algorithm~\ref{alg:main} is the application of the breadth-first-search algorithm for searching over the direction cones covering the entire ${\mathbb R}^d$. We will need some notation. As described above, let $b^{\mathbf{r}}$ be a binary sequence of length $n$ where each bit $b^{\mathbf{r}}(i)$ corresponds to a point $\mathbf{x}_i\in\{\mathbf{x}_1,\mathbf{x}_2,...,\mathbf{x}_n\}=X$ with $b^{\mathbf{r}}(i)=I(\mathbf{x}_i^\prime\mathbf{r}>0)$ for any $\mathbf{r}$ that maintains strict ordering of $\mathbf{x}\in X$ in the projection on it. Also, let $b^{0}_i$ be a zero-filled binary sequence with the $i$-th bit set to ``1'', $\oplus$ denote the binary `exclusive disjunction'$=$``XOR'' operation, $!$ be the bit inversion operator, and $\sum{b^{\mathbf{r}}}$ be the number of ``1''s in $b^{\mathbf{r}}$ (Hamming distance between $b^{\mathbf{r}}$ and $b^{0}$).

\begin{algorithm}
    \label{alg:main}
    {\bf Input:} $X=\{\mathbf{x}_1,...,\mathbf{x}_n\}\in{\mathbb R}^d$, $d<n$, $\{\mathbf{0}\}\cup X$ in general position.
    \begin{enumerate}
        \item Initialization: Calculate $X_{H_{\mathbf{x}_i}}=\{\mathbf{x}^{H_{\mathbf{x}_i}}_1,\mathbf{x}^{H_{\mathbf{x}_i}}_2,...,\mathbf{x}^{H_{\mathbf{x}_i}}_n\},i=1,...,n$, set $D=n$. Draw $\mathbf{r}_0\in S^{d-1}$ yielding a permutation $\pi_{\mathbf{r}_0}$ on ${\mathcal N}=\{1,2,...,n\}$ maintaining strict order $\mathbf{x}_{\pi_{\mathbf{r}_0}(1)}^\prime\mathbf{r}_0 < \mathbf{x}_{\pi_{\mathbf{r}_0}(2)}^\prime\mathbf{r}_0 < ... < \mathbf{x}_{\pi_{\mathbf{r}_0}(n)}^\prime\mathbf{r}_0$ and let $b^{\mathbf{r}_0}$ be the corresponding binary code. Initialize $B=\{b_1,b_2,...,b_n\}$ with $b_i=b^{\mathbf{r}_0}$ $\forall\,i=1,...,n$. Initialize a queue ${\mathcal B}_{topical}$ containing $b^{\mathbf{r}_0}$ only and an empty searchable storage ${\mathcal B}_{future}$ (e.g., binary tree).\label{step:0}
        \item For $i=1:n$ do:
        \begin{enumerate}
            \item For $j=1:n$ do:
            \begin{enumerate}
                \item If $b^{\mathbf{r}_0}(j) = 0$ then $\mathbf{x}^{H_{\mathbf{x}_i}}_j = -1 \cdot \mathbf{x}^{H_{\mathbf{x}_i}}_j$.
            \end{enumerate}
        \end{enumerate}
        \item For $i=1:\lfloor\frac{n+2}{2}\rfloor$ do:\label{step:1}
        \begin{enumerate}
            \item Pop $b$ = head of ${\mathcal B}_{topical}$, $D=\min\{D,\sum{b},n-\sum{b}\}$.\label{step:2}
            \item If $i=\lfloor\frac{n+2}{2}\rfloor$, then go to Step \ref{step:5}.
            \item For $j=1:n$ do:
            \item[] If $(b\oplus b^{\mathbf{r}_0})(j) = 0$ then
            \begin{enumerate}
                \item For $k=1:n$ do:\label{step:3}
                \begin{enumerate}
                    \item[] If $(b_j\oplus b)(k) = 1$ then $\mathbf{x}^{H_{\mathbf{x}_j}}_k = -1 \cdot \mathbf{x}^{H_{\mathbf{x}_j}}_k$, $b_j(k) = !b_j(k)$.
                \end{enumerate}
                \item If (i) $\mathbf{0}\in conv(X_{H_{\mathbf{x}_j}}\setminus\{\mathbf{0}\})$ and (ii) $(b\oplus b^0_j)\notin\mathcal{B}_{future}$
                \begin{enumerate}
                    \item[] then add $(b\oplus b^0_j)$ to $\mathcal{B}_{future}$.\label{step:4}
                \end{enumerate}
            \end{enumerate}
            \item If ${\mathcal B}_{topical}\neq\varnothing$, then go to Step \ref{step:2}, else ${\mathcal B}_{topical}={\mathcal B}_{future}$, ${\mathcal B}_{future}=\varnothing$.\label{step:5}
        \end{enumerate}
        \item {\bf Return:} $D/n$.
    \end{enumerate}
\end{algorithm}

Nontrivial is the check of condition (i) in Step \ref{step:4}, i.e. whether $\mathbf{0}\in \mbox{conv}(X_{H_{\mathbf{x}_j}}\setminus\{\mathbf{0}\})$ ($\mathbf{x}_j$ is projected into $\mathbf{0}$ in $H_{\mathbf{x}_j}$; it is excluded). In other words, given a cone $C$ unambiguously defined by the corresponding $b_j$, the linear separability (through the origin) of $X_{H_{\mathbf{x}_j},C}^+$ and $X_{H_{\mathbf{x}_j},C}^-$ has to be checked, i.e. whether $\exists$ $\mathbf{r}\in S^{d-1}\cap H_{\mathbf{x}_j}$ such that $\mathbf{r}^\prime\mathbf{x}>0$ $\forall$ $\mathbf{x}\in X_{H_{\mathbf{x}_j},C}^+$ and $\mathbf{r}^\prime\mathbf{x}<0$ $\forall$ $\mathbf{x}\in X_{H_{\mathbf{x}_j},C}^-$. This can be done by means of linear programming as follows.

Let $\mathbf{Y}$ be the $(n-1)\times(d-1)$ matrix, which rows are the points $\in(X_{H_{\mathbf{x}_j}}\setminus\{\mathbf{0}\})$ for an iteration of Step \ref{step:4} of the algorithm. The task from above narrows down to finding a feasible solution $\Lambda^0$ satisfying the constraints:
\begin{eqnarray*}
    \mathbf{Y}^\prime\Lambda &=& \mathbf{0}_{d-1},\\
    \Lambda^\prime\mathbf{1}_{n-1} &=& 1,\\
    \Lambda &\ge& \mathbf{0}_{n-1},
\end{eqnarray*}
with $\Lambda=(\lambda_1,...,\lambda_{n-1})^\prime$ and $\mathbf{0}_k$ ($\mathbf{1}_k$) being a vector-column of $k$ zeros (ones). This is what is done in the first phase of the simplex algorithm.

In Step \ref{step:0} the $X_{H_{\mathbf{x}_i}}, i=1,...,n$ --- projections of $X$ onto zero hyperplanes normal to data points $\in X$ --- are cached, an on each following step of the Algorithm for each $i$ these projections signs of several points have to be changed only, which computationally is a cheap operation. Please note, that the simplex algorithm is executed in these hyperplanes, i.e. in dimension $d-1$. This mechanism allows for further caching as well. If on Step \ref{step:4} for some $j$ $\mathbf{0}\in \mbox{conv}(X_{H_{\mathbf{x}_j}}\setminus\{\mathbf{0}\})$, a basis consisting of $d$ points will be found. If, on the next iteration of the Algorithm, on Step \ref{step:3} for the same $j$ the points changing sign do not belong to the previously found basis, clearly $\mathbf{0}\in \mbox{conv}(X_{H_{\mathbf{x}_j}}\setminus\{\mathbf{0}\})$ again, an no new execution of the simplex algorithm is needed. A more complicated caching scheme can be used here, though. One can see in Step \ref{step:1}, that the outer cycle of the Algorithm is completely deterministic and is always executed  $\lfloor\frac{n+2}{2}\rfloor$ iterations only, independent of the exact positioning of the data.

\section{Applications}\label{sec:applications}

In this section, we regard two applications of the Tukey depth: principal component analysis and supervised classification.

\subsection{Robust PCA}

First consider a problem of robust principal component analysis (PCA). \cite{Majumdar15} suggests to apply principal component analysis (PCA) to the signs of centered $X$, scaled by a depth-based distance (or ranks). Let $X^{rsgn}=\{\frac{\mathbf{x}_i - \mathbf{c}}{\|\mathbf{x}_i - \mathbf{c}\|}\bigl(D_{max} - D(\mathbf{x}_i|X)\bigr)|i=1,...,n\}$, where $D_{max}=\max_{\mathbf{x}\in\mathbb{R}^d}{D(\mathbf{x}|X)}$ and $D(\cdot|\cdot)$ is a depth. For $\mathbf{c}$ any suited estimate of center can be taken, e.g., the depth median $\mathbf{c}=\argmax_{\mathbf{x}\in\mathbb{R}^d}{D(\mathbf{x}|X)}$. Finding depth median, as well as its depth $D_{max}$ \citep{Chan04}, is a computationally nontrivial task. We obtain $X^{rsgn}$ using the depth-weighted mean $\mathbf{c}=\frac{\sum_{i=1}^n \mathbf{x}_i D(\mathbf{x}_i|X)}{\sum_{i=1}^n D(\mathbf{x}_i|X)}$ \citep[see][for the properties of the estimator]{DonohoG92}, and simply set $D_{max}$ equal to the upper bound on depth, which, for the Tukey depth, $=0.5$. Let $USV^\prime=\text{SVD}(X^{rsgn})$ be the standard singular value decomposition (SVD) of $X^{rsgn}$, where we are interested in the matrix $V$ of eigenvectors $\mathbf{v}_1,...,\mathbf{v}_d$ only. These indicate the (orthogonal) directions of the highest variance (in case this exists), or the axis of density ellipsoids.

\begin{figure}[!h]
	\begin{center}
        \includegraphics[keepaspectratio=true,scale=0.515]{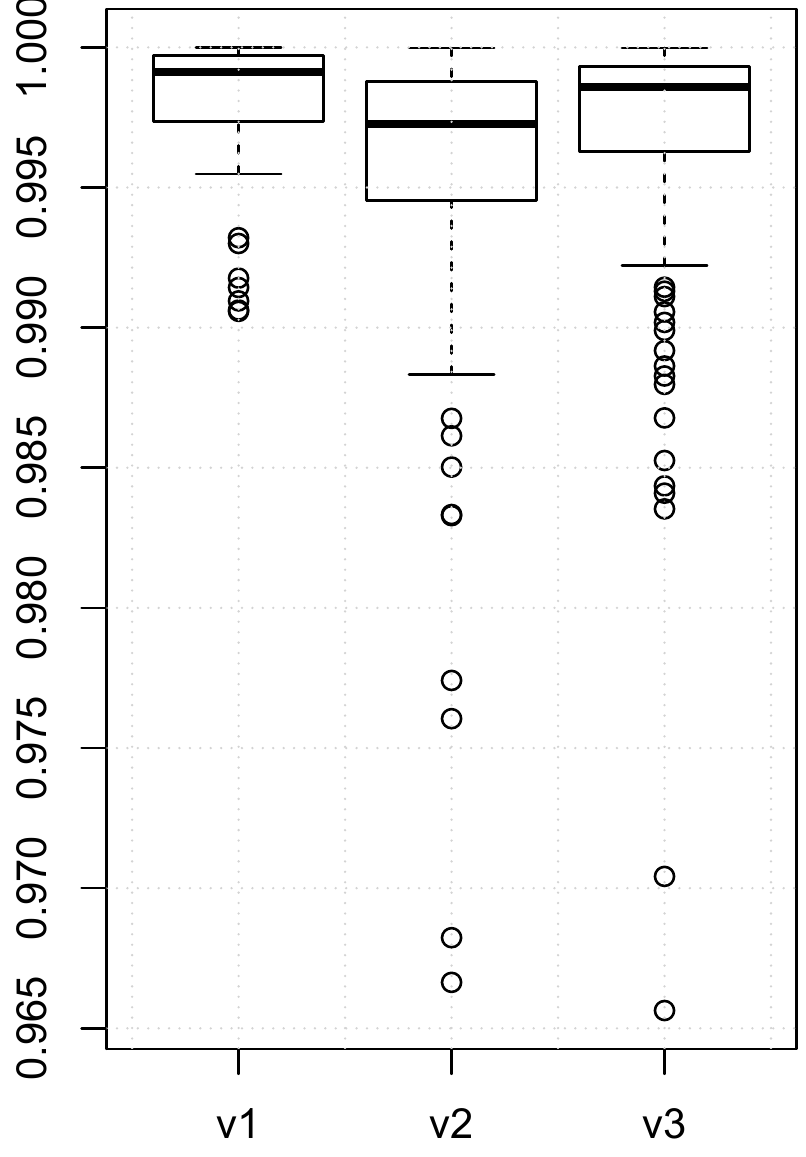}\quad\includegraphics[keepaspectratio=true,scale=0.515]{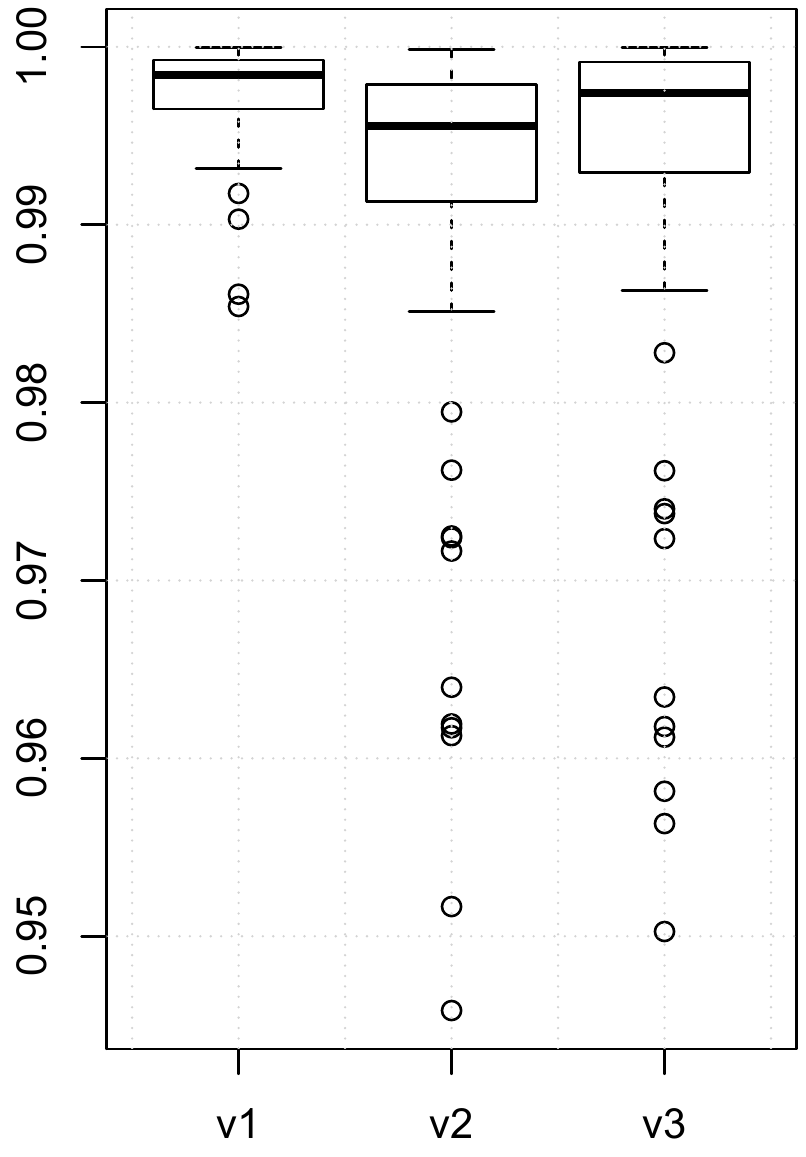}\quad\includegraphics[keepaspectratio=true,scale=0.515]{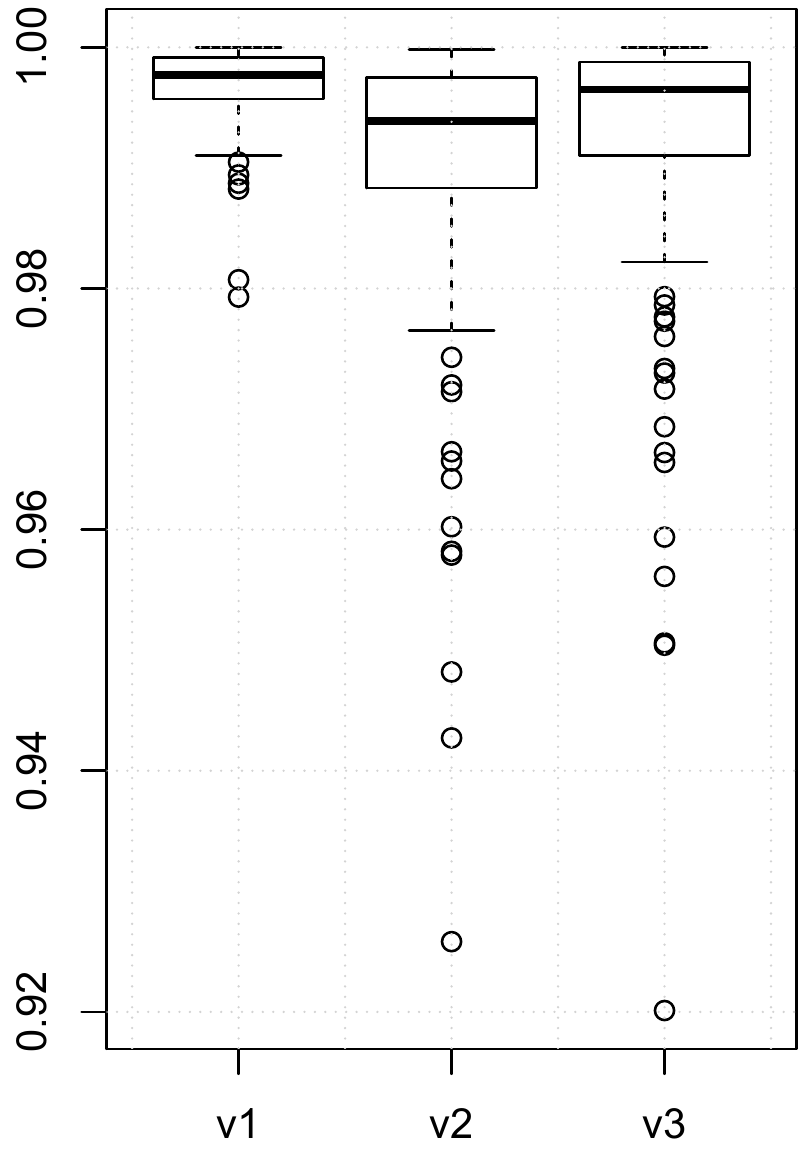}
	\end{center}
	\caption{Boxplots of the inner products of obtained eigenvectors with the true ones for 100 points drawn from the Gaussian elliptically symmetric distribution centered in $\mu$ and scattered as $\Sigma$ over 100 runs; for the standard SVD algorithm (left), ROBPCA (middle), and the depth-based PCA (right).}
	\label{fig:svdNormal}
\end{figure}

\begin{figure}[!h]
    \begin{center}
        \includegraphics[keepaspectratio=true,scale=0.515]{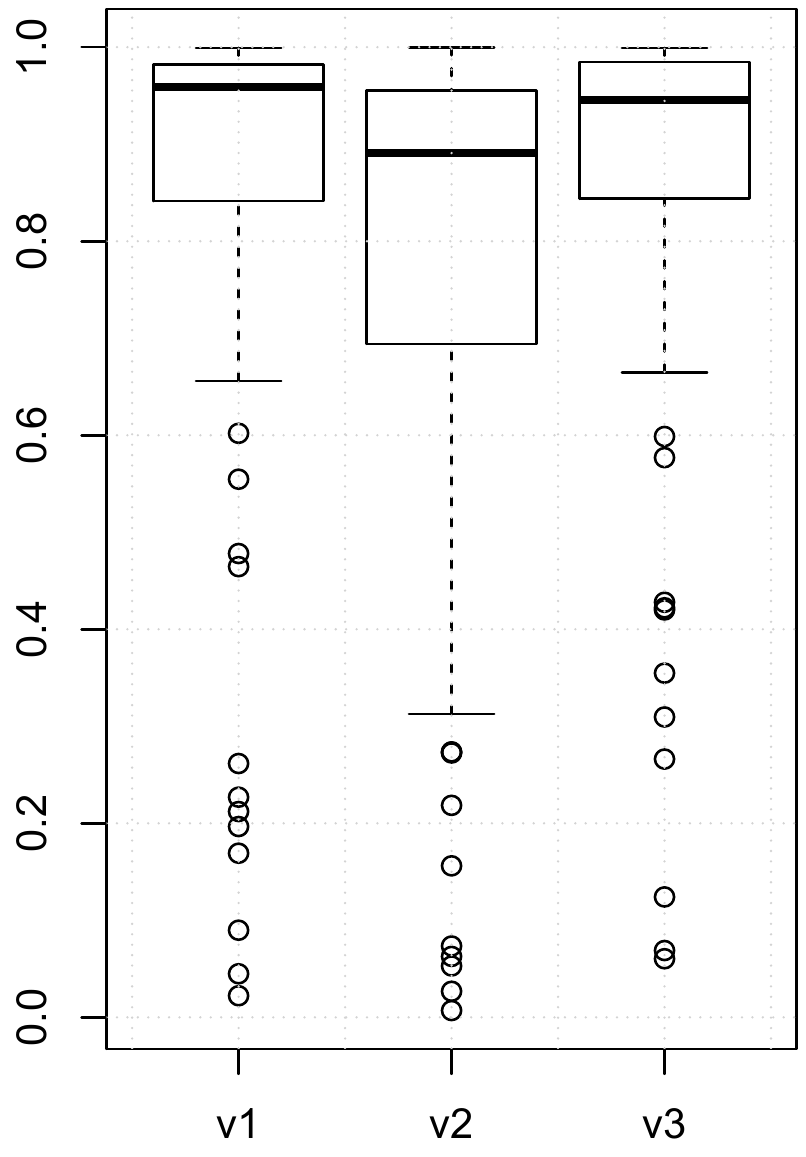}\quad\includegraphics[keepaspectratio=true,scale=0.515]{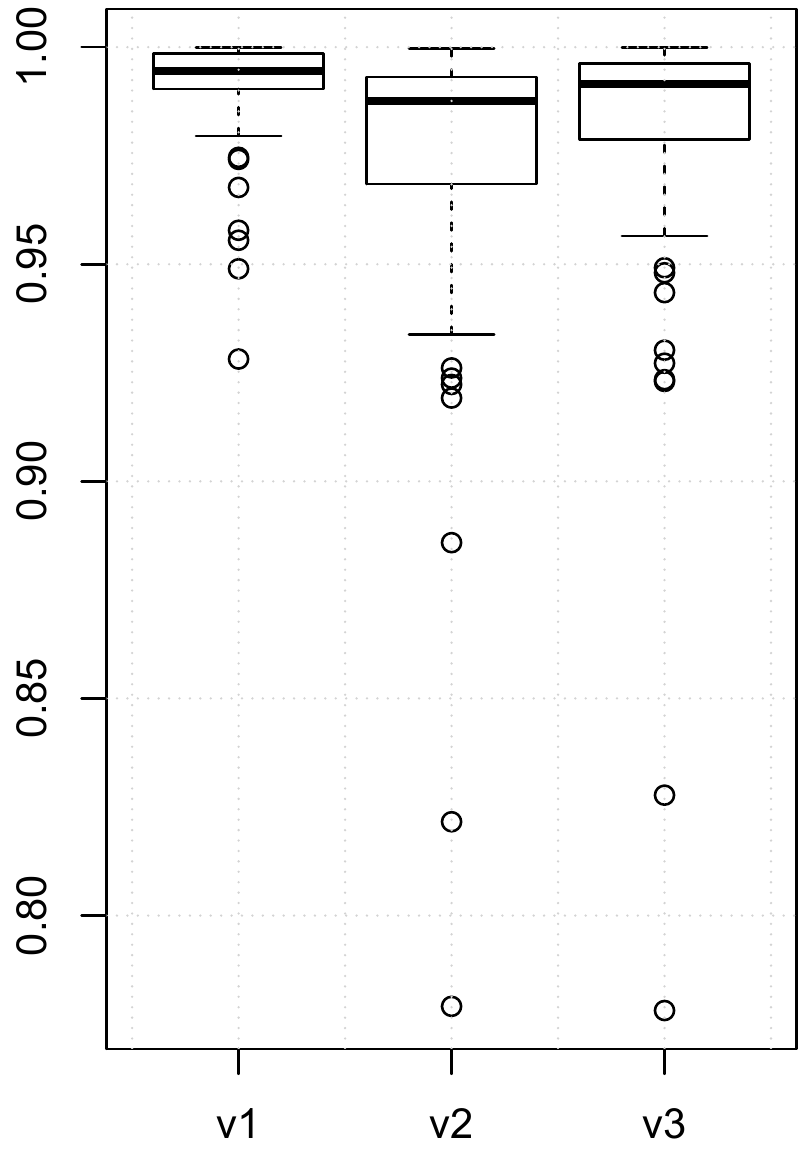}\quad\includegraphics[keepaspectratio=true,scale=0.515]{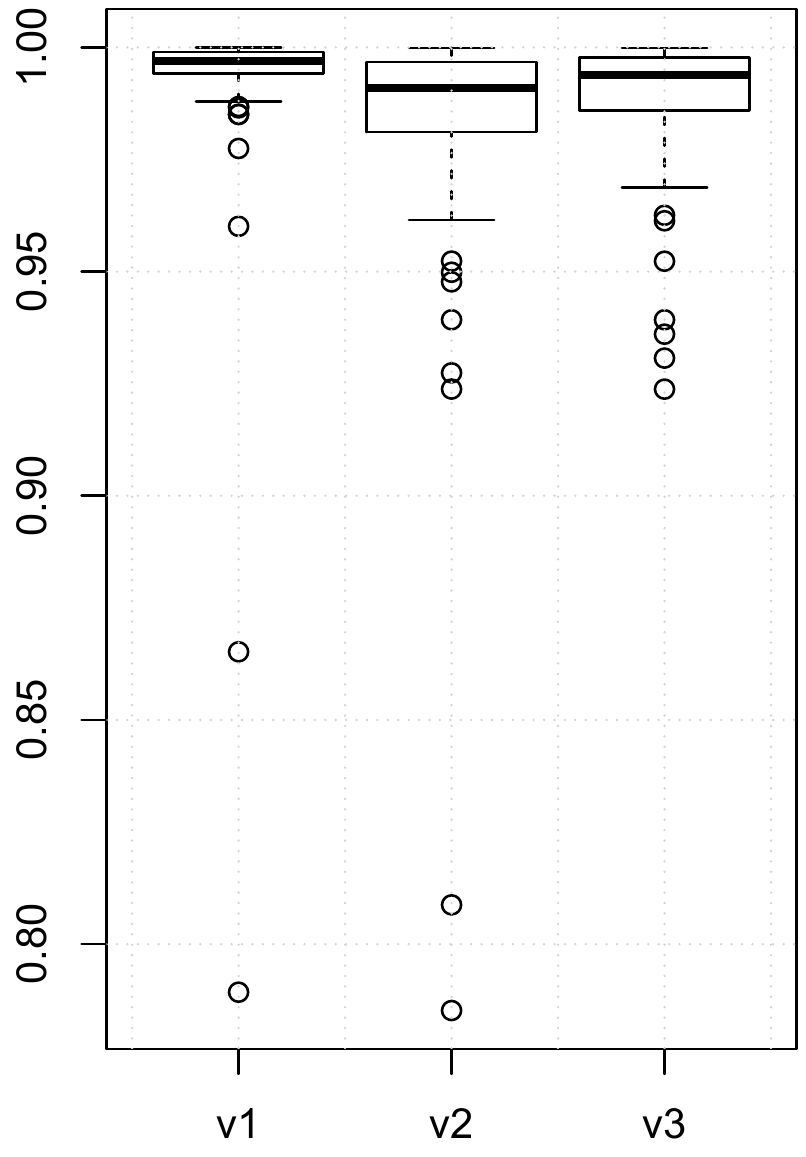}
    \end{center}
	\caption{Boxplots of the inner products of obtained eigenvectors with the true ones for 100 points drawn from the Student-$t_1$ (Cauchy) elliptically symmetric distribution centered in $\mu$ and scattered as $\Sigma$ over 100 runs; for the standard SVD algorithm (left), ROBPCA (middle), and the depth-based PCA (right).}
	\label{fig:svdCauchy}
\end{figure}

We contrast this method based on the exactly computed Tukey depth with the standard SVD algorithm and with one of the most powerful contemporary tools for the robust PCA, the ROBPCA method by \cite{HubertRVB05}, on a 3-dimensional elliptically symmetric Gaussian and Student-$t_1$ (Cauchy) distribution with $\boldsymbol{\mu}=(1, 1, 1)^\prime$ and structure matrix $\Sigma=\begin{pmatrix} 1 & 1 & 1 \\ 1 & 4 & 4 \\ 1 & 4 & 10 \end{pmatrix}$. Setting $n=100$ we repeat the experiment 100 times for each setting. Boxplots of the inner products of the obtained eigenvectors with the true ones are presented in Figures~\ref{fig:svdNormal} and~\ref{fig:svdCauchy} for Gaussian and Student $t_1$ distributions respectively. (Ideally they all should be equal to one.)


For the Gaussian case, although one can observe that the standard SVD outperforms the other two methods (price of robustness), the loss of performance is not dramatic. When switching to the Cauchy setting though, the standard method performs plainly wrong: its inner products cover the entire range between 1 and 0 where the obtained eigenvectors are orthogonal the the wished ones. While both other methods behave quite well, the depth-based PCA outperforms the ROBPCA slightly, which can be explained by the following. The ROBPCA relies much on the Minimum Covariance Determinant \citep[MCD, see][]{Rousseeuw84} which assumes that there is a portion of data preserving its elliptical shape, while the Tukey depth describes data geometry in a non-parametric way.

\subsection{Robust supervised classification}

\begin{figure}[!h]
	\begin{center}
        \includegraphics[keepaspectratio=true,scale=0.615]{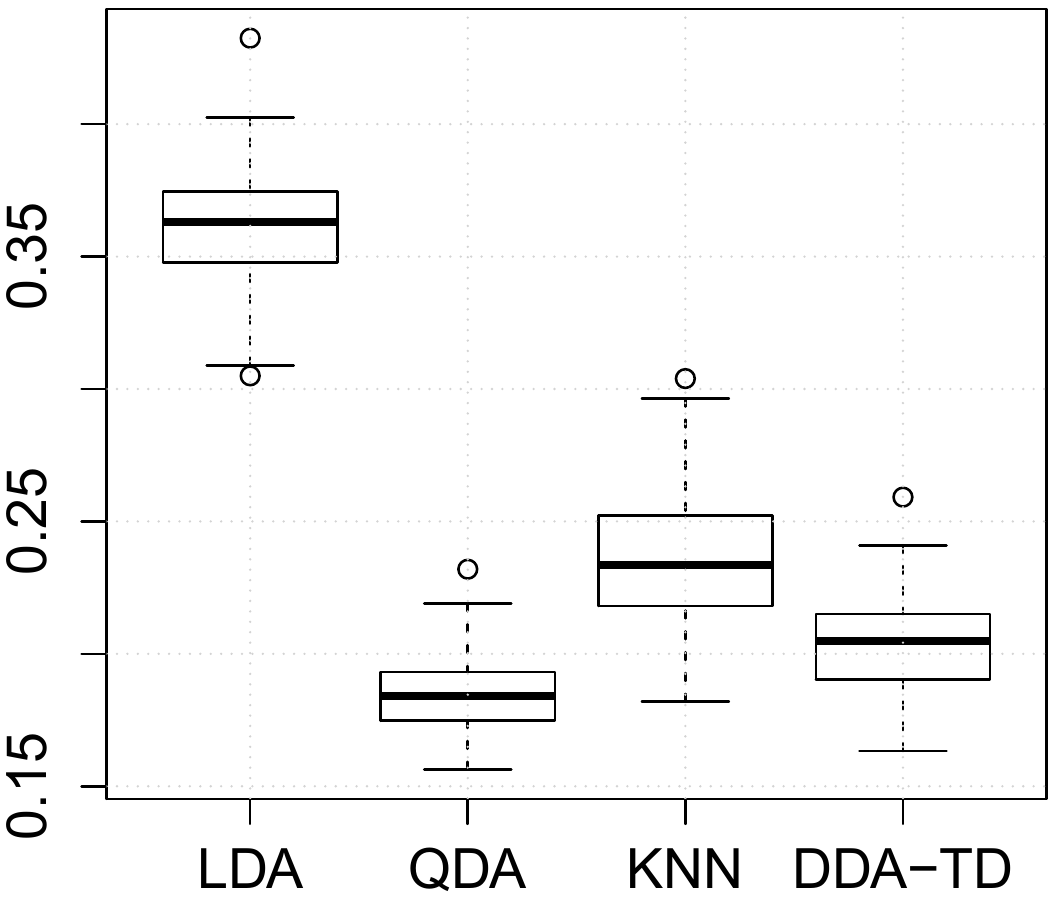}\quad\includegraphics[keepaspectratio=true,scale=0.615]{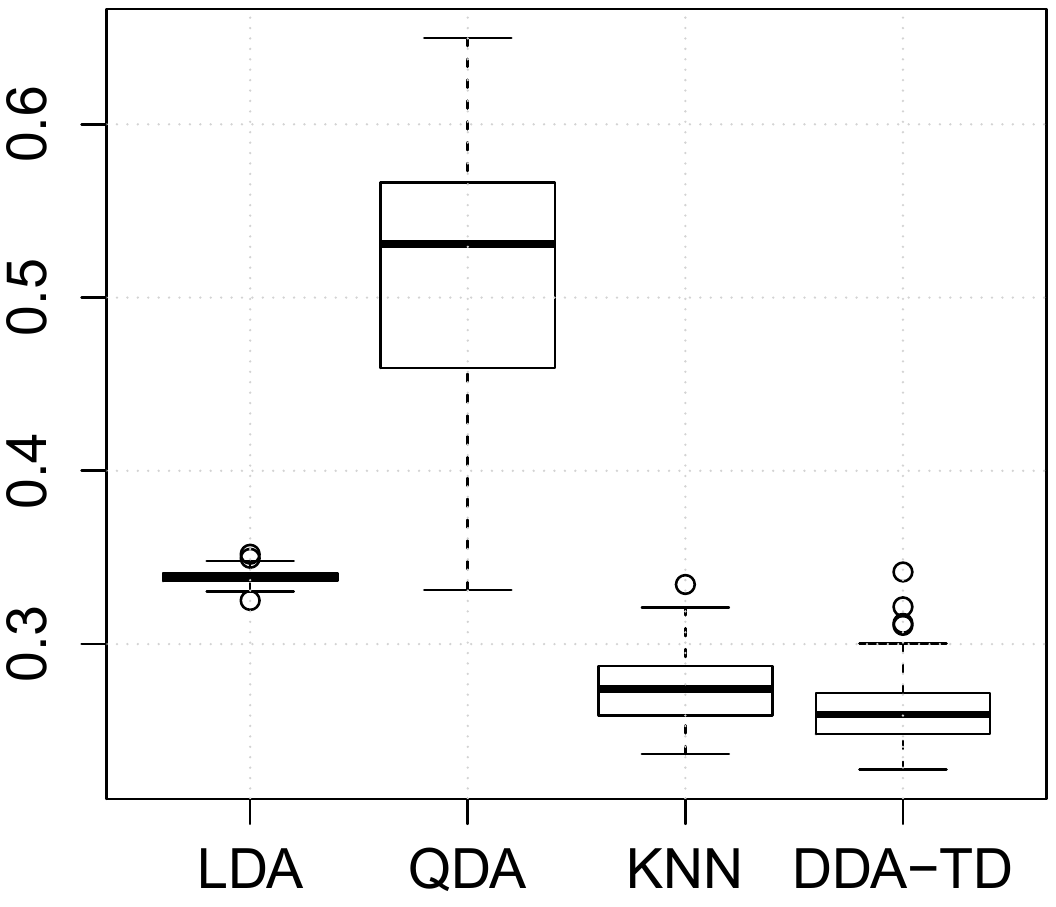}
    \end{center}
	\caption{Boxplots of the error rates over 100 tries for two classes generated from elliptically symmetric Gaussian (left) and Student-$t_1$ (right) distributions.}
	\label{fig:dda}
\end{figure}

\citet{LiCAL12} introduced a depth-based classification principle exploiting the idea of the depth-vs-depth plot ($DD$-plot). Given $X_1=\{\mathbf{x}_1,...,\mathbf{x}_{n_1}\}$ and $X_2=\{\mathbf{x}_{n_1+1},...,\mathbf{x}_{n_1+n_2}\}$ two training classes in $\mathbb{R}^d$, the $DD$-plot is a subset of the unit square $[0,1]^2$ $X^{DD}=\{(D(\mathbf{x}_i|X_1),D(\mathbf{x}_i|X_2))|i=1,...,n_1+n_2\}$. Any classifier can be applied to $X^{DD}$. We employ the $DD\alpha$-classifier --- a fast heuristic technique based on the $\alpha$-procedure \citep{VasilevL98}; the reader is referred to \citet{LangeMM14} for the detailed reference. Here we demonstrate the performance of the $DD\alpha$-classifier with the exactly computed Tukey depth compared to three standard techniques: Linear Discriminant Analysis (LDA), Quadratic Discriminant Analysis (QDA), and $k$-nearest neighbors classifiers (KNN) for a pair of elliptically distributed classes differing in location, scale, shape and prior, a particularly complicated case. Taking $\Sigma$ from above as a basis, we generate $X_1$ and $X_2$ each from Gaussian or Student-$t_1$ elliptically symmetric distributions with parameters $\pi_1=\frac{2}{3}$, $\mu_1=(0, 0, 0)^\prime$, $\Sigma_1=4\times\begin{pmatrix} 1 & 1 & 1 \\ 1 & 4 & 4 \\ 1 & 4 & 10 \end{pmatrix}$ , respectively $\pi_2=\frac{1}{3}$, $\mu_2=(1, 1, 1)^\prime$, $\Sigma_2=\begin{pmatrix} 4 & 1 & 4 \\ 1 & 1 & 1 \\ 4 & 1 & 10 \end{pmatrix}$. We generate in total 200 training and 1000 testing points and exclude outsiders \citep[points lying beyond the convex hull of each of the classes, see][for additional information]{MozharovskyiML15} when calculating the error rate. Boxplots of the corresponding error rates over 100 tries are shown in Figure~\ref{fig:dda}.

In general the $DD\alpha$-classifier coupled with the exactly computed Tukey depth performs well, for the Gaussian distribution its is outperformed by the QDA only, which is optimal in this setting. For the Student-$t_1$ distribution it outperforms such a strong competitor as KNN. Regarding the Student-$t_1$ elliptical setting, it is worth to mention that a further study not directly related to the current presentation and thus skipped here, shows that using the $DD\alpha$-classifier with other depths such as projection depth or Mahalanobis depth with robust estimates of location and scatter, allows to achieve performance similar to this when using the Tukey depth and even better.

\section{Conclusions}\label{sec:outlook}

The paper presents an algorithm computing the exact Tukey depth by finding a global minimum over a finite range of variants. The task of computing the Tukey depth is NP-complete while all separations of $X$ into two subsets by hyperplanes through $\mathbf{z}$ are regarded. The algorithm follows the traditions of the cone segmentation of a finite-dimensional space and regards candidate cones of constant depth according to a first-breadth order. It employs the initial idea of~\citet{LiuZ14a}, but identifies a facet using linear programming and exploits the fact that each point $\in X$ changes the halfspace only once during the entire execution of the breadth-first search algorithm. Linear programming is executed in $\mathbb{R}^{d-1}$ and the found basis can be cached for each of these $n$ $(d - 1)$-dimensional projections. Also, binary coding of the cones does not require computation of their location.

The fact that the task of computation of the Tukey depth is of non-polynomial time complexity, a number of existing procedures, deterministic and output-sensitive ones, but also different platforms and often absence of implementation intricate the speed comparison, or even make this unreasonable. Because of this, in the current presentation a possible simulation study is substituted by application examples. In general, one can expect that the computation time of exact algorithms grows exponentially with increasing dimension. For this reason, in practical tasks the Tukey depth is often approximated, see e.g. \citet{Dyckerhoff04}, \citet{CuestaAlbertosNR08}, or \citet{ChenMW13}.

Nevertheless exact algorithms are needed to assess the approximating ones. On the other hand, if the dimension is kept fixed, the time complexity is polynomial. Considering the task of the supervised classification from Section~\ref{sec:connections}, it is the number of points that increases and not the dimension. As another example, one can regard multivariate functional halfspace depth \citep{ClaeskensHSV14}, where even for a (high-dimensional) functional data set the number of output dimensions can still be quite low, which allows for exact computation of time marginals for their subsequent integration.

\section*{Acknowledgements}
Major part of the research has been conducted during the PhD study of the author at the University of Cologne.
The author is grateful to the supervisor of his PhD thesis professor Karl Mosler for his valuable comments on the earlier version of the paper
and to his colleagues professor Rainer Dyckerhoff and Pavel Bazovkin for their helpful remarks.
The author wants to thank Xiaohui Liu for providing the Matlab source code of the algorithm from the work by \cite{LiuZ14a} and for his insightful comments on the earlier version of the manuscript.



\end{document}